%Paper: hep-th/9305071
%From: palev@fwet.rug.ac.be (Palev)
%Date: Sun, 16 May 93 10:42:29 +0100

\baselineskip=16pt

\noindent
\hskip 4in Int. Report TWI-93-17
\vskip 18pt
{\bf A SUPERALGEBRA MORPHISM OF $U_q[OSP(1/2N)]$ ONTO THE DEFORMED
OSCILLATOR SUPERALGEBRA $W_q(N)$}

\vskip 18pt
\noindent
T. D. Palev\footnote*{Permanent address: Institute for Nuclear Research
and Nuclear Energy, 1784 Sofia, Bulgaria; E-mail
palev@bgearn.bitnet}

\noindent
Applied Mathematics and Computer Science, University of Ghent,
B-9000 Gent, Belgium

\noindent
and

\noindent
International Centre for Theoretical Physics, 34100 Trieste, Italy

\vskip 24pt
\leftskip 32pt
{\bf Abstract.} We prove that the deformed oscillator superalgebra
$W_q(n)$ (which in the Fock representation is generated essentially
by $n$ pairs of $q$-bosons) is a factor algebra of the quantized
universal enveloping algebra $U_q[osp(1/2n)]$.
We write down a $q$-analog of the Cartan-Weyl basis for the
deformed $osp(1/2n)$ and give also an oscillator realization of all
Cartan-Weyl generators.


\vskip 32pt
\leftskip 12pt
{\bf I.Introduction}

\vskip 24pt
In Refs.1,2  a realization of the quantized
universal enveloping algebra $U_q\equiv U_q[osp(1/2n)]$ of the
orthosymplectic
Lie superalgebra $osp(1/2n)$ has been constructed
in terms of $n$ pairs of independent
(mutually commuting) $q$ oscillators $b_i^\pm, i=1,\ldots,n $, i.e.,
in terms of  $q$-deformed Bose creation and annihilation operators
as introduced in [3-5]. In other words it has been shown that there
exists an algebra morphism $\pi$ of $U_q[osp(1/2n)]$ into the deformed
Weyl (or oscillator) algebra $W_q(n)$. The main purpose of the present
note is to show that $\pi$ is in fact a morphism of $U_q[osp(1/2n)]$
onto $W_q(n)$ in the category of associative superalgebras. The latter
implies that $W_q(n)$ is a factor algebra of $U_q[osp(1/2n)]$. In this
way we solve the conjecture, formulated in [6], and give a partial
answer to the more general hypothesis [7], namely that there exists
a morphism of  $U_q[osp(2m+1/2n)]$ onto the Clifford-Weyl superalgebra
$W_q(m/n)$, generated by $m$ pairs of $q$-fermions and $n$ pairs of
$q$-bosons. The case of $U_q[osp(1/2)]$   was studied in Refs. 8 and 9 in
relation to the deformed para-Bose operators. In Ref. 6 it was proved
for $U_q[osp(1/4)]$. The same case was considered also in Ref.
10  in relation to
the "supersingleton" Fock representation of $U_q[osp(1/4)]$ and its
singleton [11] structure.

We proceed  to recall the definitions of $W_q(n)$   and of
$U_q[osp(1/2n)]$ in the notation we are going to use. Following
[12], we consider $W_q(n)$ as an associative algebra with unity 1, free
generators $b_i^\pm,\; k_i, \; k_i^{-1}$  and  relations
($i,j=1,\ldots ,n$)
$$k_i^{-1}k_i=k_ik_i^{-1}=1$$
$$k_ib_i^\pm=q^{\pm 1} b_i^\pm k_i$$
$$b_i^-b_i^+-q^2b_i^+b_i^-=k_i^{-2} \eqno(1)$$
$$b_i^-b_i^+-q^{-2}b_i^+b_i^-=k_i^2$$
$$a_ia_j=a_ja_i,  \hskip 12pt  i\neq j,$$
where $a_i=b_i^\pm, k_i^{\pm 1}$.
To turn $W_q(n)$ into a superalgebra (  {\bf Z}$_2$-graded algebra) we set

 $$deg(b_i^\pm)=1 \in {\bf Z}_2 , \hskip 12pt deg(k_i^{\pm 1})=0
 \in {\bf Z}_2, \hskip 12pt  i=1,\ldots,n. \eqno(2)$$

In the Fock representation of $W_q(n)$ , namely when $k_i=q^{N_i}$,
$b_i^\pm$ are
the deformed $q$-bosons and $N_i$ is the $i$th boson number operator [3-5].
With respect to the grading induced from (2) $W_q(n)$ is an
infinite-dimensional associative superalgebra.

The superalgebra $U_q[osp(1/2n)]$ can be defined in different
equivalent ways [1,2,13]. We choose the Cartan matrix $(\alpha_{ij})$
as in [13], i.e., this is a $n$-by-$n$ symmetric matrix with

$$\alpha_{nn}=1, \; \alpha_{ii}=2, \; i=1,\ldots,n-1, \;
\alpha_{j,j+1}=\alpha_{j+1,j}=-1, \; j=1,\ldots,n-1, \eqno(3)$$
and all other $\alpha_{ij}=0$ . Then $U_q[osp(1/2n)]$ is the free
associative superalgebra with Chevalley generators
$E_i,\; F_i,\; K_i,\; i=1,\ldots,n, $ graded as

$$deg(E_n)=deg(F_n)=1 \in {\bf Z}_2, \;
deg(E_i)=deg(F_i)=deg(K_j)=0 \in {\bf Z}_2,\;i=1,\ldots,n-1,\;
j=1,\ldots,n, \eqno(4)  $$
which satisfy the Cartan relations

$$K_iK_i^{-1}=K_i^{-1}K_i=1, \quad K_iK_j=K_jK_i, \quad
 i,j=1,\ldots,n, \eqno(5) $$

$$K_iE_j=q^{\alpha_{ij}}E_jK_i, \quad
K_iF_j=q^{-\alpha_{ij}}F_jK_i, \quad  i,j=1,\ldots,n, \eqno(6)  $$

$$[E_n,F_n]={{K_n^2-K_n^{-2}}\over{q-q^{-1}}}, \quad
[E_i,F_j]=\delta_{ij}{{K_i^2-K_i^{-2}}\over{q^2-q^{-2}}}
\quad \forall \; i,j=1,\ldots,n \; \; {\rm except} \;i=j=n,
\eqno(7) $$
the Serre relations for the simple positive root vectors

$$ [E_i,E_j]=0, \quad if \quad i,j=1,\ldots,n \quad
and \quad \vert i-j \vert >1,
\eqno(8)  $$

$$ E_i^2E_{i+1}-(q^2+q^{-2})E_iE_{i+1}E_i+E_{i+1}E_i^2=0,
   \quad i=1,\ldots,n-1, \eqno(9)  $$

$$ E_i^2E_{i-1}-(q^2+q^{-2})E_iE_{i-1}E_i+E_{i-1}E_i^2=0,
   \quad i=2,\ldots,n-1, \eqno(10)  $$

$$E_n^3E_{n-1}+(1-q^2-q^{-2})(E_n^2E_{n-1}E_n+E_nE_{n-1}E_n^2)+
E_{n-1}E_n^3=0, \eqno(11)  $$
and the Serre relations obtained from (8)-(11) by replacing
everywhere $E_i$ by $F_i$.

Here and throughout the paper

$$[a,b]=ab-(-1)^{deg(a)deg(b)}ba, \eqno(12)  $$

$$[a,b]_{q^n}=ab-(-1)^{deg(a)deg(b)}q^nba, \eqno(13)  $$
for any two homogeneous elements $a,b \in U_q$
and it is assumed that the deformation parameter $q$ is any
complex number except $q=0$, $q^2=1$ and $q^4=1$.

The action of the coproduct $\Delta :U_q \rightarrow U_q \otimes
U_q$, antipode $S:U_q \rightarrow U_q$ and counit $\varepsilon :
 U_q \rightarrow {\bf C}$ can be given, for instance, as:

$$\Delta (E_i)=E_i \otimes K_i + K^{-1} \otimes E_i,\quad
\Delta (F_i)=F_i \otimes K_i +K_i^{-1}\otimes F_i ,\quad
\Delta(K_i)=K_i\otimes K_i, \eqno(14)   $$

$$S(E_i)=-q^{\alpha_{ii}}E_i, \quad
S(F_i)=-q^{-\alpha_{ii}}F_i, \quad
S(K_i)=K_i^{-1}, \eqno(15)   $$

$$\varepsilon(E_i)= \varepsilon(F_i) = \varepsilon(K_i) =0,
\quad \varepsilon(1)=1. \eqno(16)$$

PROPOSITION 1. The linear map $\pi :U_q[osp(1/2n)]
\rightarrow W_q(n)$, given as
$$\vcenter {\openup2\jot \halign{$#$ \hfil
   & \hskip 24pt $#$ \hfill &  \hskip 24pt $#$ \hfill \cr
\pi(E_i)=-b_i^-b_{i+1}^+, & \pi(F_i)=-b_{i+1}^-b_i^+,
 & \pi(K_i)=k_i^{-1}k_{i+1},  \cr
\pi(E_n)=-b_n^-, & \pi(F_n)=b_n^+, & \pi(K_n)=q^{-{1 \over 2}}k_n \cr
}}\eqno(17) $$
and extended by associativity
defines a morphism {}   (in the category of the associative
superalgebras) of $U_q[osp(1/2n)]$ into the oscillator
superalgebra  $W_q(n)$.

The proof is an immediate consequence of the results of [2]
and the observation that the map $\rho : W_q(n) \rightarrow
W_q(n) $

$$b_i^+ \rightarrow \xi b_i^-, \quad b_i^- \rightarrow -\xi
b_i^+,  \quad
k_i \rightarrow q^{-1}k_i^{-1}, \quad \xi =1 \; or -1,\;
i=1,\ldots,n \eqno(18) $$
defines an authomorphism in $W_q(n)$.

We wish to show that $\pi$ is a map of $U_q[osp(1/2n)]$ onto
$W_q(n)$. To this end consider the following $2n$ elements from
$U_q[osp(1/2n)] \; (i=1,\ldots, n-1)$:

$$\vcenter{\openup3\jot\halign {$#$ \hfil \cr
B_i^-=-\sqrt{2q \over {(q+q^{-1})}}
[E_i,[E_{i+1},[E_{i+2},[\ldots,[E_{n-2},[E_{n-1},E_n]_{q^{-2}}
]_{q^{-2}}\ldots ]_{q^{-2}}K_iK_{i+1} \ldots K_n, \cr
B_n^-=-\sqrt{2q \over {(q+q^{-1})}}E_nK_n \cr
}} \eqno(19)   $$
\vskip 12pt
\noindent
and
$$\vcenter{\openup3\jot\halign {$#$ \hfil \cr
B_i^+=\sqrt{2q \over {(q+q^{-1})}}
[\ldots,[F_n,F_{n-1}]_{q^2},F_{n-2}]_{q^2},\ldots]_{q^2},
F_{i+2}]_{q^2},F_{i+1}]_{q^2},F_{i}]_{q^2}
K_i^{-1}K_{i+1}^{-1} \ldots K_n^{-1}, \cr
B_n^+=\sqrt{2q \over {(q+q^{-1})}}F_nK_n^{-1}. \cr
}} \eqno(20)   $$

{}From (17), (19) and (20) one derives expresions for the
$q$-oscillator realization of $B_1^{\pm},\ldots,B_n^{\pm},$
$$\pi (B_i^-)=q^{-2(n-i)}\sqrt{2 \over {(q+q^{-1})}}
b_i^-k_i^{-1}k_{i+1}^{-2}k_{i+2}^{-2}\ldots k_{n}^{-2} ,\quad
i=1,\ldots,n, \eqno(21)   $$

$$\pi (B_i^+)=q^{2(n-i)+1}\sqrt{2 \over {(q+q^{-1})}}
b_i^+k_ik_{i+1}^{2}k_{i+2}^{2}\ldots k_{n}^{2} ,\quad
i=1,\ldots,n. \eqno(22)   $$

At $q=1$ the images $\pi(B_i^{\pm})$  of $B_i^{\pm}$
reduce to the usual
nondeformed Bose creation and annihilation operators,
whereas the nondeformed $B_1^{\pm},\ldots,B_n^{\pm}$ satisfy the
trilinear relations $( \{x,y\} \equiv xy+yx ) $

$$[\{B_i^\xi,B_j^\eta\} ,B_k^\epsilon]=
(\epsilon -\xi)\delta_{ik}B_j^\eta +
(\epsilon-\eta)\delta_{jk}B_i^\xi,\quad i,j,k=1,\ldots,n,\quad
\xi, \eta, \epsilon =\pm \; or \pm 1. \eqno(23)   $$

\noindent
In quantum field theory the operators (23) are known as
para-Bose operators. They were introduced by
Green [14] as a possible generalization of the Bose statistics.
The para-Bose operators are a good substitute for the Chevalley
generators in the sense that they also define completely the Lie
superalgebra $osp(1/2n)$ [15]. More precisely, consider the
operators

$$\vcenter{\openup5\jot \halign{ \hskip 98pt $#$ \hfil & \hskip 26pt $#$
\hfil & \hskip 84pt $#$ \hfil \cr
B_i^\xi, & i=1,\ldots,n, & (24a) \cr
H_i=-{1\over 2} \{B_i^+,B_i^-\}  , & i=1,\ldots,n, & (24b) \cr
{1\over 2} \{B_j^+,B_k^-\} , & j \not=k=1,\ldots,n, & (24c) \cr
{1\over 2} \{B_p^\xi,B_q^\xi\}, &
p\leq q=1,\ldots,n, \quad \xi= \pm \; or \pm1. & (24d) \cr
}}$$

\noindent
Then the operators $H_i$ constitute a basis in a Cartan
subalgebra $H$ of $osp(1/2n)$, the para-Bose operators $(24a)$
-- a basis in the odd subspace of $osp(1/2n)$, all
anticommutators
$(24b-d)$ -- a basis in the even subalgebra $sp(2n)$ and
$(24b,c)$ -- a basis in the subalgebra $gl(n)$. The
operators $(24a,c,d)$ are root vectors (of $H$) and all
operators (24) give (one possible) Cartan-Weyl basis in $osp(1/2n)$.
Replacing in (24) the para-Bose operators with Bose operators,
one obtains the usual ladder realizations of $osp(1/2n)$,
$sp(2n)$ and $gl(n)$.

The deformed operators $B_1^{\pm},\ldots,B_n^{\pm},$ which are
in fact deformed para-Bose operators, possess similar properties.
Elsewhere we shall show that these operators together with the "Cartan"
elements $K_1^{\pm 1},\ldots,K_n^{\pm 1}$ define uniquely
$U_q[osp(1/2n)]$ and hence give an alternative definition of this
quantum algebra  in terms of generators, which have
more immediate physical significance (see for more discussions along
this line Ref.6). Also without a proof we formulate a proposition,
which solves in a simple way the problem
 for constructing a $q-$analog of the Cartan-Weyl
basis in the deformed $osp(1/2n)$ superalgebra.

\vskip 6pt
PROPOSITION 2. The $q$-analog of the Cartan-Weyl basis of
$osp(1/2n)$ is given with the set of all operators as follows:

$$\vcenter{\openup5\jot \halign{ \hskip 100pt $#$ \hfil & \hskip 26pt $#$
\hfil & \hskip 144pt $#$ \hfil \cr
K_i^{\pm}, & i=1,\ldots,n, & (25a) \cr
B_i^\pm, & i=1,\ldots,n, & (25b) \cr
{1\over 2}\{B_j^+,B_k^-\} , & j \not=k=1,\ldots,n, & (25c) \cr
{1\over 2}\{B_p^\xi,B_q^\xi\} , &
p < q=1,\ldots,n, \quad \xi= \pm \ & (25d) \cr
(B_r^\pm)^2, & r=1,\ldots,n. & (25e) \cr
}}$$

\noindent
The set of all ordered monomials (with respect to a certain
normal order [13]) of the operators
$(25a-d)$ constitute a basis in $U_q[osp(1/2n)]$.

Remark 1. The operators $(25a,c)$ appear as a $q$-analog of the
Cartan-Weyl basis of the deformed $gl(n)$. These operators
generate a Hopf subalgebra of $U_q[osp(1/2n)]$, which is
isomorphic to $U_q[gl(n)]$. The latter follows from the
observation that the generators and the relations of $U_q[gl(n)]$
are among the generators and the relations of $U_q[osp(1/2n)]$.

Remark 2. The operators $(25a,c-d)$ generate a subalgebra of
$U_q[osp(1/2n)]$, which is a deformation of the universal
enveloping algebra of $sp(2n)$. This subalgebra, however, is not
a Hopf subalgebra of $U_q[osp(1/2n)]$ even in the simplest case of
$n=1$ [16].

The $q$-oscillator realization of $(25a,b)$ was already written
(see eqs.(17), (21) and (22)). The realization of the rest of
the Cartan-Weyl generators reads:

$$\vcenter{\openup5\jot \halign{ \hskip 33pt $#$ \hfil & \hskip 20pt $#$
\hfil & \hskip 30pt $#$ \hfil \cr
\pi(\{B_i^-,B_j^+\})=2q^{2(i-j)}b_i^-b_j^+k_i^{-1}k_{i+1}^{-2}\ldots
k_{j-2}^{-2}k_{j-1}^{-2}k_j^{-1}, & i<j=1,\ldots,n, & (26) \cr
\pi(\{B_i^-,B_j^+\})=2q^{2(i-j)}b_i^-b_j^+k_jk_{j+1}^{2}\ldots
k_{i-2}^{2}k_{i-1}^{2}k_i, & i>j=1,\ldots,n, & (27) \cr
\pi(\{B_i^-,B_j^-\})=2q^{2(i+j-2n)+1}b_i^-b_j^-k_i^{-1}k_j^{-1}
k_{i+1}^{-2}\ldots k_n^{-2}
k_{j+1}^{-2} \ldots k_n^{-2}, & i\neq j=1,\ldots,n, & (28) \cr
\pi(\{B_i^+,B_j^+\})=2q^{2(2n-i-j)+3}b_i^+b_j^+k_ik_j
k_{i+1}^{2}\ldots k_n^{2}
k_{j+1}^{2} \ldots k_n^{2}, & i\neq j=1,\ldots,n. & (29) \cr
}}$$
\vskip 12pt
PROPOSITION 3. The oscillator algebra $W_q(n)$ is a factor
algebra of $U_q \equiv U_q[osp(1/2n)] $ (in the sense of
associative superalgebras).

Proof. For a proof it sufficies to show that the generators of $W_q(n)$,
namely $b_i^\pm, \; k_i^{\pm 1}, \; i=1,\ldots,n$ are among the images of
$U_q$ under the map $\pi$ (see proposition 1).

Keeping in mind that $\pi$ is a linear map, which preserves the
multiplication, i.e., $\pi(ab)=\pi(a)\pi(b)$ and as a consequence
also $\pi(c^{-1})= (\pi(c))^{-1}$, one derives from
$\pi(K_i)=k_i^{-1}k_{i+1}, \;i=1,\ldots,n-1$
 and $ \pi(K_n)=q^{-{1 \over 2}}k_n$ [see (17)] that

$$k_i=\pi(q^{-{1\over 2}}K_i^{-1}K_{i+1}^{-1} \ldots K_n^{-1}),
\quad k_i^{-1}=\pi(q^{1\over 2}K_iK_{i+1} \ldots K_n),
\quad i=1,\ldots,n \eqno(30)   $$
and therefore
$$k_i, k_i^{-1} \in \pi(U_q), \quad i=1,,\ldots,n. \eqno(31)  $$

{}From (21) we have that for each $i=1,\ldots,n$

$$b_i^{-}=q^{2(n-i)}\sqrt{2\over {q+q^{-1}}}\pi(B_i^-)
       k_ik_{i+1}^2k_{i+2}^2\ldots k_n^2.$$

\noindent
Inserting here the expressions for $\pi(B_i^-)$ and $k_i$ from (19) and
(30) we obtain after some rearrangement of the order of the generators

$$b_i^-=\pi(-q^{n-i}[E_i,[E_{i+1},[E_{i+2},\ldots
[E_{n-2},[E_{n-1},E_n]_{q^{-2}} \ldots ]_{q^{-2}}
\prod _{k=1}^{n-i}K_{i+k}^{-2k}), \quad i=1,\ldots,n-1. \eqno(32)   $$

\noindent
In a similar way one has

$$b_i^+=\pi(q^{i-n}[\ldots[F_n,F_{n-1}]_{q^2},F_{n-2}]_{q^2},
\ldots ]_{q^2},F_{i+2}]_{q^2},F_{i+1}]_{q^2},F_i]_{q^2}
\prod _{k=1}^{n-i}K_{i+k}^{2k}), \quad i=1,\ldots,n-1. \eqno(33) $$

\noindent
{}From (17), (31), (32) and (33) we conclude that

$$k_i, k_i^{-1}, b_i^\pm \in \pi(U_q), \quad i=1,\ldots,n, \eqno(34)  $$

\noindent
which completes the proof.

In a forthcoming paper we will discuss in more details the description
of $U_q \equiv U_q[osp(1/2n)] $ in terms of
$B_i^{\pm} ,\; K_i^{\pm 1},\; i=1,\ldots,n $. Here we first  fix the
terminology.

\vskip 6pt
DEFINITION. The generators $B_i^{\pm} ,\; K_i^{\pm 1},\; i=1,\ldots,n $
of $U_q$ will be called pre-oscillator generators. The description of
$U_q[osp(1/2n)] $ (or of any subalgebra $A \subset U_q$)
in terms of the pre-oscillator generators will be
refered to as pre-oscillator or pre-ladder form
(realization, description) of $U_q$
(resp. of $A$).

The pre-oscillator form of $U_q$ gives an alternative description to the
usual Chevalley realization, where all elements of $U_q$ are functions
of $E_i,\, F_i, \, K_i^{\pm 1}, \; i=1,\ldots,n.$ The name
pre-oscillator
comes to remind that the image of any subalgebra $A \subset U_q$
under the map $\pi$ [see (17)],
$$\pi:\; A \rightarrow W_q(n), \eqno(35)$$
gives a ladder (or oscillator) realization of $A$.

The pre-oscillator form of the subalgebra $A$, which need
not necessarily be a Hopf
subalgebra of $U_q$, can be used for construction of representations
of $A$ in any tensorial power of Fock spaces $F_q(n)$ of the oscillator
algebra $W_q(n)$. This stems from the observation that the
action of the comultiplication $\Delta$ is well defined on the
pre-oscillator generators, whereas it is not
defined on their $\pi-$images, i.e., on the deformed
$q-$bosons. In order to be more concrete, define by induction
a morphism

$$\Delta ^{(k)}=[(\otimes_{i=1}^{k-2}id)\otimes \Delta]\circ
\Delta^{(k-1)}, \hskip 12pt \Delta^{(2)}=\Delta, \hskip 12pt
\Delta^{1}=id. \eqno(36) $$

\noindent
of $U_q$ into $\otimes_{i=1}^{k}U_q$. The map

$$\pi^{(k)}=(\otimes_{i=1}^{k}\pi) \circ \Delta ^{(k)}:
U_q[osp(1/2n)] \rightarrow \otimes_{i=1}^{k}W_q(n) \eqno(37)  $$

\noindent
is a morphism of $U_q[osp(1/2n)] $ into $\otimes_{i=1}^{k}W_q(n)$.
Consider $W_q(n)$ in its Fock space $F_q(n)$ representation,
$W_q(n) \subset End[F_q(n)]$. Then $\pi^{(k)}(A)$ gives a representation
of $A$ in $\otimes_{i=1}^{k}F_q(n)$. In particular if $A$ is given in
a pre-oscillator form, then for any
$a=a(B_1^{\pm},\ldots, B_n^{\pm},K_1^{\pm 1},\ldots,
 K_n^{\pm 1} ) \in A  $ the map
$\pi^{(k)}$, defined on each element $a
\in A$ as

$$\pi^{(k)}(a)=a(\pi^{(k)}B_1^{\pm},\ldots,
 \pi^{(k)}B_n^{\pm},\pi^{(k)}K_1^{\pm 1},\ldots,
 \pi^{(k)}K_n^{\pm 1} ), \eqno(38)  $$

\noindent
is a representation of $A$ in $\otimes_{i=1}^k F_q(n)$. Certainly the
same statement holds if $A$ were given in a Chevalley form. The point
is however that in the practical applications $A$ is usually given
directly in terms of $q-$deformed bosons. From this realization it
is fairly evident how to reconstruct the pre-oscillator form of $A$, to
which one can apply the above technique in order to consctruct tensor
products of Fock representations.  The main difficulty will be
then to decompose the tensor product $\otimes_{i=1}^{k}F_q(n)$,
considered as an $A-$module, into a direct sum of irreducible
(and, may be, also indecomposible)  $A-$modules.

\vskip 24pt
{\bf Acknowledgements}
\vskip 12pt
The author is thankful to Prof. Abdus Salam for the kind
hospitality at the International Center for Theoretical Physics,
Trieste. He is grateful to
Prof. Vanden Berghe for the possibility to visit the
Department of Applied Mathematics and Computer Science at the
University of Ghent, where the paper was completed.
It is a pleasure to thank Dr. J. Van der Jeugt and Dr. N. I.
Stoilova for stimulating discussions. The research was
supported through  contract $\Phi - 215$ of
the Committee of Science of Bulgaria.

\vskip 24pt
{\bf References}
\vskip 12pt

\noindent
1. Floreanini, R.,Spiridonov, V.P. and Vinet, Phys.lett.
{\bf B234}, 383 (1990).

\noindent
2. Floreanini, R.,Spiridonov, V.P. and Vinet, L., Comm.Math.Phys.
{\bf 137}, 149 (1991).

\noindent
3. Biedenharn, L.C.,J.Phys. A {\bf 22}, L873 (1989).

\noindent
4. Macfarlane, A.J.,J.Phys. A {\bf 22}, 4581 (1989).

\noindent
5. Sun, C.P. and Fu, H.C.,J.Phys. A {\bf 22}, L983 (1989).

\noindent
6. Palev, T.D. and Stoilova, N.I., Preprint ICTP IC/93/54 (1993)
(to appear in Lett.Math.Phys.).

\noindent
7. Palev, T.D., Preprint Concordia Univ. 5/92 (1992) (to appear in
Journ.Math.Phys.).

\noindent
8. Floreanini, R. and Vinet, L., Preprint UCLA/90/TEP/30;
    J.Phys. A {\bf 23}, L1019 (1990).

\noindent
9. Celeghini, E., Palev, T.D. and Tarlini,  M., Preprint
 YITP/K-879, Kyoto (1990); Mod.Phys.Lett. B

  \hskip -2pt
{\bf 5}, 187 (1991).

\noindent
10. Flato, M.,  Hadjiivanov, L.K. and Todorov, I.T., Quantum
Deformations of Singletons and of Free

\hskip -2pt Zero-mass Fields,
to appear in Foundations of Physics, the volume dedicatet to A. O. Barut.

\noindent
11. Flato, M. and and Fronsdal, C., Singletons: Fundamental Gauge Theory,
in: {\bf Topological and

\hskip -2pt Geometrical Methods in Field Theory},
Symposium in Espoo, Finland 1986, Eds. J. Hietarinta,

\hskip -2pt J. Westerholm
(World Scientific, Singapore 1986) pp. 273-290.

\noindent
12. Hayashi, T., Comm.Math.Phys. {\bf 127}, 129 (1990).

\noindent
13. Khoroshkin, S.M. and Tolstoy, V.N., Comm.Math.Phys, {\bf 141}, 599 (1991).

\noindent
14. Green, H.S., Phys.Rev. {\bf 90} , 270 (1953).

\noindent
15. Ganchev, A. and Palev, T.D., J.Math.Phys. {\bf 21}, 797 (1980).

\noindent
16. Kulish, P.P. and Reshetikhin, N.Yu., Lett.Math.Phys. {bf
18}, 143 (1989).
\end